# Gene–Environment Interplay in the Social Sciences

Rita Dias Pereira, Pietro Biroli, Titus Galama, Stephanie von Hinke, Hans van Kippersluis, Cornelius A. Rietveld, and Kevin Thom

**Summary**

Nature (one's genes) and nurture (one's environment) jointly contribute to the formation and evolution of health and human capital over the life cycle. This complex interplay between genes and environment can be estimated and quantified using genetic information readily available in a growing number of social science data sets. Using genetic data to improve our understanding of individual decision making, inequality, and to guide public policy is possible and promising, but requires a grounding in essential genetic terminology, knowledge of the literature in economics and social-science genetics, and a careful discussion of the policy implications and prospects of the use of genetic data in the social sciences and economics.

# Introduction

In the Sisyphean debate on the role of "nature" (i.e., one's genes) versus "nurture" (i.e., one's environment) in explaining life outcomes, economists have traditionally sided with the importance of environmental influences (Smith, 1776).[1] However, over time, the role of genetic variation in shaping life outcomes has become increasingly accepted.[2] Early instances of this interest explored the genetic roots of human behavior (Becker, 1976; Hirshleifer, 1977) and equality of opportunity (Rawls, 1971; see also Harden, 2021b; Pereira, 2021) as well as the intergenerational transmission of human capital (Behrman & Taubman, 1976; Bowles & Nelson, 1974), where "genetic talents" are considered one of the sources of inequality, or of the intergenerational persistence in human capital.[3] More recently, economists have sought to quantify the role of genes on educational and health outcomes (see, e.g., Barth et al., 2020; Miller et al., 1995). Economists increasingly acknowledge the importance of genetic differences between individuals, considering them intimately connected with decision-making processes and capabilities that ultimately explain variation in individuals' health and economic outcomes over the life course (see, e.g., Beauchamp et al., 2011; Benjamin et al., 2012, 2020; Fletcher, 2011).

Genetic effects do not operate in isolation, however. The traditional notion that "nature" and "nurture" operate independently is now considered to be obsolete (Plomin et al., 1977; Rutter, 2006; Turkheimer, 2000). Indeed, it is argued that pitting "nature" *against* "nurture" should be relinquished in favor of a more systemic view that considers the complex interplay that exists between people's genetic variation and the environment in which they develop (Haldane, 1946; Harden, 2021a; Heckman, 2007; Hunter, 2005; Leibowitz, 1974; Rutter, 2006). With recent advances in genetic discovery and the growing availability of genetic data in many social science data sets, it has become possible to quantify and estimate the importance of gene–environment interplay, incorporating the effects of genes and environments, as well as their interaction.

This article presents an overview of studies in the economics and social science literature that examine gene–environment interplay on outcomes of interest to economists. The overview includes studies that explore not only the main genetic and environmental effects, but also whether and how one's genetic variation can moderate the effect of environmental circumstances, or, vice versa, how one's environment can modify the genetic effects on individuals' outcomes of interest. We focus on recent papers that use molecular measures of genetic differences—either single genetic variants or indices that aggregate large numbers of (typically very small) influences of genetic variants, so-called polygenic indices (PGIs), or polygenic scores (PGSs)—and specific measures of environmental exposures, preferably predetermined or exogenous. The purpose of the review is not to be complete or exhaustive, but rather to show the breadth and the potential of this type of research for economists and social scientists who are interested in studying and understanding the origin and the evolution of inequality in health and human capital.

The remainder of this article is structured as follows. A short introduction to genetics, genome-wide association studies, and polygenic indices is followed by a review of the literature on gene–environment interplay. A discussion of potential policy implications, as well as different avenues for future research, follows.

## Introduction to Human Genetics

This brief introduction to human genetics provides the necessary vocabulary and understanding of the main genetic concepts that are used in this research area. For a more in-depth discussion, see, for example, Mills et al. (2020).

Human DNA consists of a sequence of about three billion pairs of nucleotide molecules. These nucleotides come in four varieties (or "letters"): adenine (A), guanine (G), cytosine (C), and thymine (T), where "A" always pairs with "T" and "C" always pairs with "G." Certain sections (stretches of nucleotides) of the DNA that code for proteins are called genes. It is estimated that humans have approximately 25,000 genes that code for proteins with a specific function. All nucleotides together constitute the human genome, which is divided into 23 pairs of chromosomes (22 autosomal pairs and the pair of sex chromosomes). Within each pair, one chromosome is inherited from the biological mother, and one from the biological father.

Most (99.9%) human DNA is identical between unrelated individuals. This means that most DNA codes for characteristics are common to humans. Locations (or "loci") where DNA varies between individuals are called polymorphisms. Polymorphisms code for differences between individuals and are often the object of interest in genetic studies. The most common form of a polymorphism is a *Single Nucleotide Polymorphism* (SNP, pronounced "snip"), a change at a single nucleotide locus. Figure 1 shows this graphically, where two individuals are shown with identical DNA, apart from one (highlighted) nucleotide; a SNP. These two variants of the nucleotides are called alleles, and the less common allele in a given population is called the minor allele. For a polymorphism to be studied, there needs to be sufficient variation; a commonly used threshold is a minor allele frequency (i.e., the frequency of the least common allele) of at least 1% (i.e., at least 1% of the population carries the least common variant). In the human genome, there are approximately 85 million SNPs with a minor allele frequency of less than 1% (1000 Genomes Project Consortium et al., 2015). Less common types of genetic variants are small-scale insertions/deletions (*indels*) and *structural variants* (Feuk et al., 2006; Weischenfeldt et al., 2013). Indels refer to insertion and/or deletion of nucleotides in genomic DNA of less than 1 kilobase (kb) in length; structural variants are usually longer and comprise changes of nucleotides into genomic DNA larger than 1 kb, such as deletions, copy number variants, duplications, insertions, inversions, and translocations (Feuk et al., 2006). While SNPs are most commonly used in social science genetics to construct indices of genetic predisposition, indels and structural variants are more commonly used in medical studies or candidate gene studies (we discuss the latter in the section "Candidate Gene Studies").

<COMP: INSERT FIGURE 1 NEAR HERE>

In 2003, the Human Genome Project completed the first successful sequencing of the entire human genome. This scientific breakthrough allowed for the direct measurement of all nucleotides that constitute the DNA code. The subsequent advent of inexpensive genotyping chips opened the door to the collection of genetic information in several social science data sets, which allow researchers to identify linkages between a person's genetic variants and important life outcomes such as health and socioeconomic status (Benjamin et al., 2012; Visscher et al., 2017).

## *Genome-Wide Association Studies*

The most common genetic variation is of the SNP kind and is what is most commonly used in social science genetic studies. Identifying specific SNPs that are associated with a particular outcome is done through so-called genome-wide association studies (GWASs). A GWAS relates each of the $J$ SNPs (denoted by $x_{ij}$, where $i$ refers to the individual and $j = 1, \ldots, J$) to the outcome of interest $y_i$ in a hypothesis-free approach. These regressions ideally include all SNPs simultaneously, but since data sets have fewer individuals than SNPs, such a model is not identified (Benjamin et al., 2012). GWASs therefore evaluate millions of independent regressions, one for each SNP $j$, as:

$$y_i = \beta_j^{GWAS} x_{ij} + z'_i \gamma + \varepsilon_{ij} \quad (1)$$

where $x_{ij}$ is coded as 0, 1, or 2, reflecting the number of minor alleles carried by individual $i$, the vector $z_i$ denotes the relevant controls, and $\varepsilon_{ij}$ is the error term.[4] This produces a list of $\hat{\beta}_j^{GWAS}$ estimates for all $J$ SNPs.

To account for multiple hypothesis testing as well as the fact that SNPs that are closely spaced on the genome are likely to be co-inherited (so-called "Linkage Disequilibrium"), GWASs use very stringent significance levels, with a statistically significant effect typically identified as those with $p < 5 \times 10^{-8}$. This stringent significance level, in combination with the tiny effect sizes of individual SNPs (see, e.g., Chabris et al., 2015; Rietveld et al., 2013) requires extremely large samples to ensure adequate power. For example, the very first GWAS of educational attainment used a sample of 125,000 people and identified three genome-wide significant loci (Rietveld et al., 2013). The second GWAS included 400,000 participants and identified 74 genome-wide significant loci (Okbay et al., 2016), the third GWAS used data on 1.1 million individuals, identifying 1,271 lead-SNPs associated with educational attainment (Lee et al., 2018), and the fourth GWAS of educational attainment used 3 million individuals and identified 3,952 lead-SNPs (Okbay et al., 2022).

## *Aggregation Into an Index: Polygenic Indices*

Finding a large number of genome-wide significant SNPs is typical for virtually all outcomes of interest to social science researchers. Virtually all human traits are highly "polygenic" (Chabris et al., 2015; Visscher et al., 2008). That is, there is no "gene for" these outcomes; instead, the vast majority of human outcomes are affected by many SNPs, each with very small effect sizes.[5] To maximize the amount of variation in a outcome that is explained by genetic variation, we can create a measure of genetic risk or predisposition for each individual that reflects the aggregation of numerous small contributions of millions of genetic loci into a unidimensional index. Such summary measures of genetic predisposition have been called polygenic indices (PGIs) or polygenic scores (PGSs) and are generally defined as a weighted sum of SNPs, reflecting the best linear genetic predictor of an outcome (e.g., Mills et al., 2020). In its most basic form, a PGI is constructed as:

$$PGI_i = \sum_{j=1}^{J} \hat{\beta}_j^{GWAS} x_{ij}, \qquad (2)$$

where $x_{ij}$ again denotes the number of copies of the minor allele for individual $i$ at SNP $j$, and $\hat{\beta}_j^{GWAS}$ are the $\beta$ coefficients from Equation 1 (Dudbridge, 2013). Weighting each SNP by its effect size ensures that SNPs with large effect sizes are weighted more heavily than those with small effect sizes.[6] The predictive power of a polygenic index also depends on the GWAS sample size, since the $\hat{\beta}_j^{GWAS}$ estimates become more precise with larger samples, reducing measurement error. For example, the polygenic index based on the first GWAS of educational attainment explains 3%–4% of the variation in educational attainment out-of-sample (Rietveld et al., 2014). The polygenic index based on the second GWAS explains 6%–8% (Okbay et al., 2016), the polygenic index based on the third GWAS explains 11%–13% (Lee et al., 2018), and the polygenic index based on the fourth GWAS explains 12%–16% (Okbay et al., 2022).

The maximum explained variance of a polygenic index is related to the outcome's SNP-based heritability, which is the share of the variance of an outcome that can be explained using SNPs as an input in a variance-decomposition method. Using methods like genome-based restricted maximum likelihood (GREML) estimation (Yang et al., 2011), studies have shown that the SNP-based heritability of educational attainment is 25% (Rietveld et al., 2013). Hence, polygenic indices for educational attainment based on the most recent (third) GWAS explain about half the variation in educational attainment that is related to individual genetic variants. The discrepancy between these two measures of the explanatory power of genetic variation has been dubbed the "hidden heritability" (Tropf et al., 2017). It has been shown that the predictive power of many polygenic indices is still substantially smaller than the SNP-based heritability estimates (Davies et al., 2016; Okbay et al., 2016; Tropf et al., 2017; Visscher et al., 2008). One of the reasons for this is measurement error in the polygenic index (Becker et al., 2021), which can be addressed using an instrumental variable approach (see, e.g., DiPrete et al., 2018; Van Kippersluis et al., 2021).

## *Interpretation of Genetic Associations*

To avoid the temptation of interpreting these genetic associations as biologically determined, it is fundamental to remember that the estimated associations between the genetic variants and the outcomes are specific to the environmental and demographic context of the GWAS discovery sample (Benjamin et al., 2020; Domingue et al., 2020; Mills et al., 2020). As such, their effects should not be interpreted as solely biological, genetic, or immutable (e.g., Mostafavi et al., 2017; Kweon et al., 2020); the effects depend on the environmental and cultural context. For example, some obesity-related SNPs may not be captured in a GWAS that includes individuals who do not have easy access to calorie-dense foods.

Furthermore, polygenic indices should not be given a causal interpretation. Although they are fixed at conception and therefore are predetermined, they are generally correlated with the environments shaped by parents and other ancestors. This phenomenon, called "gene–environment correlation" (*rGE*), refers to the fact that certain environments are more prevalent for carriers of certain genotypes (Fletcher & Conley, 2013; Plomin et al., 1977). For example, Biroli and Zünd (2021) show that individuals with a high polygenic index for alcohol consumption are more likely to live near (and move to) areas characterized by a higher density of pubs.

A more subtle form of environmental confounding of polygenic indices is that they capture the environments that are shaped by one's parents. Indeed, if children have a high polygenic index for a particular outcome, parents will also be more likely to have a high polygenic index for that same outcome. It has been shown that parental genotype influences offspring outcomes through genetic *as well as* environmental pathways. For example, parents who smoke will pass on their "smoking genes," but also their "smoking environment." Similarly, parents with a high polygenic index for educational attainment will not only pass their "education genes" to their children, but may also differentially shape the rearing environment compared to parents with a low polygenic index for educational attainment. In fact, Kong et al. (2018) shows that the polygenic index for educational attainment constructed from SNPs that were *not* transmitted to children remains strongly predictive of children's educational attainment. Since these SNPs were not transmitted, this correlation has to be driven by an environmental channel. The phenomenon where parental genes influence offspring outcomes through environmental pathways is known as "genetic nurture."

To get closer to a causal interpretation of the effects of these genetic variants, accounting for such genetic nurture, one can perform a within-family analysis. A family fixed effect will hold constant the effects associated with maternal or paternal genotype.[7] Indeed, within-family

(Muslimova et al., 2020; Selzam et al., 2019) and adoptee (Cheesman et al., 2020) studies have shown that the predictive power of polygenic indices is reduced when the rearing environment provided by biological parents is statistically controlled for. For behavioral and social phenotypes like education or childbearing, the within-family effect is about half the size of the between-family effect.

A related but distinct and broader idea is that, in many cases, genes act through the environment (e.g., Cheesman et al., 2020; Mostafavi et al., 2020). This means that associations between genes and certain outcomes are dependent on the environment, and therefore genetic studies can inform us only about specific populations of people living in a given time and space. Therefore, changes in the associations between genes and outcomes over time and space can yield interesting insights into how the social, economical, cultural, and historical context alters the influence of genetics on outcomes.

In summary, a polygenic index is a predictive and useful measure of an individual's genetic propensity towards an outcome, but its effect should be interpreted with care. For a detailed discussion of the interpretation of the estimated coefficients in a gene–environment specification, including settings with exogenous and endogenous measures for genes and the environment, see Biroli et al. (2021).

# Review of the Gene–Environment Interplay Literature

In this section, we discuss some of the recent literature in economics and the social sciences that has focused on the estimation of gene–environment interplay. Given the recent advances in human genetics described in the previous section, we focus on papers that use *molecular* measures of genetic differences, instead of studies that exploit family designs without molecular measures, such as twin, sibling, or adoptee-studies (for an overview of such approaches, see, for example, Briley et al., 2015; Kohler et al., 2011).[8]

But which molecular measures of genetic differences should be used in gene–environment interplay studies: individual genetic variants or polygenic indices? The answer to this question is not immediately obvious. Studying the interplay between the environment and single genetic variants allows one to narrow down the underlying biological mechanisms that may be driving the genetic effect, as laboratory studies or animal models can experiment with the genetic variation under study. However, studies based on single variants might lack statistical power. This is the main reason why most recent studies use polygenic indices, where statistical power is generally less of an issue. However, polygenic indices do not come without limitations. Importantly, the biological mechanisms captured by these polygenic indices are often poorly or only partially understood.

For the purpose of this article, this section begins by discussing the initial spread and the limitations of studies that leverage so-called "candidate genes"—genetic variants that were handpicked by researchers based on a priori knowledge of the gene's biological function and a hypothesized association with a particular outcome. It then discusses the SNP-environment interplay literature, focusing on associations between economic outcomes and some of the "top SNPs," i.e., single polymorphisms identified in large GWASs whose associations with the outcomes have been consistently replicated and whose effects are found to be large (i.e., an explanatory power $R^2 \geq 0.1\%$). Studies considered here use endogenous, predetermined, as well as exogenous environments. The section then turns to a discussion of the literature on gene–environment interplay that uses polygenic indices. Given the ever-growing size of this literature, the discussion focuses on studies that use environments that are predetermined or exogenous, since these allow for a more straightforward interpretation of the estimated gene-by-environment interplay (Biroli et al., 2021).

## *Candidate Gene Studies*

Candidate gene studies started appearing in the early 2000s. These studies typically started from a biologically based hypothesis that suggested that a specific genetic variant would have an impact on an outcome (e.g., mutations in the adiponectin gene were expected to be linked with type 2 diabetes, as the gene produces adiponectin, which was known to modulate the metabolism of lipid and glucose; Kondo et al., 2002). In 2009, over 34,000 published articles reported gene–disease associations in about 1,500 journals (Little et al., 2009). However, as early as 2011, researchers began to question the validity and replicability of candidate gene studies (Duncan & Keller, 2011), with direct implications for gene–environment studies. The main critique is that candidate gene studies were statistically underpowered and resulted in false positive associations between genetic variants and the outcomes of interest (Chabris et al., 2012; Duncan et al., 2014; Harden, 2021a; Hewitt, 2012; Okbay & Rietveld, 2015; Sullivan, 2007). Indeed, we now know that the effect sizes of individual genetic variants are generally well below an $R^2$ of 0.1% (Harden, 2021a). Indeed, candidate gene studies of complex outcomes have rarely been replicated (e.g., Border et al., 2019; Duncan & Keller, 2011; Ioannidis, 2005; Karlsson Linnér et al., 2019). This lack of replication led the journal *Behavior Genetics* to issue an editorial policy in 2012, stating that novel hypotheses for candidate gene studies would be considered for publication only if they met the statistical criteria of genome-wide significance, taking into account all sources of multiple testing. Alternatively, the novel hypothesis would have to be sufficiently well-powered and accompanied by a replication. Studies conducting a replication of former studies, or a meta-analysis referring to the same genetic variant, would also be considered. Following these guidelines, later and better-powered publications did not find support in favor of numerous previously reported associations between outcomes and candidate-genes such as MAOA, 5-HTTLPR, COMT, DRD2, and DAT1 (likely the most extensively studied candidate genes for substance use and externalizing disorder; Samek et al., 2016). Additionally, a large ($N = 62,138$ to $N = 443,264$) preregistered analysis focusing on 18 highly studied candidate genes for depression concluded that there was "no support for historical candidate gene or candidate gene-by-interaction hypotheses for major depression across multiple large samples" (Border et al., 2019). Plenty of other studies followed, finding a lack of evidence supporting the importance of other candidate-genes (e.g., between the oxytocin receptor gene OXTR and antisocial behavior; Poore & Waldman, 2020). Caution is therefore warranted when reviewing the literature on gene–environment interplay based on genetic variants identified in candidate gene studies.

### *Single Nucleotide Polymorphism–Environment Interplay*

After the robust and replicable identification of associations between SNPs and outcomes using the hypothesis-free GWAS approach, the field started to use these genome-wide significant SNPs in the study of gene–environment interplay. Two commonly studied outcomes in this literature are Body Mass Index (BMI, a measure of obesity) and smoking.

*Body Mass Index*

The connection between SNPs located in the FTO gene and obesity was first documented by one of the first very influential GWASs (Frayling et al., 2007). This result was replicated in 13 different cohorts totalling 38,759 individuals. Each additional copy of the rs9939609[9] allele was associated with an average BMI increase of 0.10 Z-score units, corresponding to an $R^2$ of 1%. This finding has since been replicated multiple times in many different cohorts (see, e.g., Babenko et al., 2019; Dina et al., 2007), including in the Han Chinese population (Chiang et al., 2019), ethnic Mongolians (Zhang et al., 2018), Brazilians (Da Fonseca et al., 2019), and Polish men (Sobalska-Kwapis et al., 2017). The discovery of the connection between the FTO gene and obesity can be considered one of the success stories of GWASs (Visscher et al., 2017), which also prompted detailed studies of the biological functioning of this gene (see, e.g., Claussnitzer et al., 2015; Smemo et al., 2014).

The association between *FTO* gene and BMI and obesity is widely accepted, and there have been many studies of *FTO*-by-environment interplay. Summarizing past successes of gene–environment interaction research, Ritz et al. (2017) argue that success depends on the availability of high-quality exposure assessment and longitudinal measures, study populations with a wide range of exposure levels, the inclusion of ethnically and geographically diverse populations, and, last but not least, large population samples. The authors highlight the discovery of an interaction between the *FTO* gene and physical activity as one of the successes in this line of research.

A study on 17,808 Danes found that the association between *FTO* and body fat mass and body fat accumulation is moderated by physical activity, such that a weaker relationship is found for the group most physically active (Andreasen et al., 2008). A year later, this finding was replicated for BMI and waist circumference in a sample of 20,371 participants (Vimaleswaran et al., 2009). Another study conducted on an Indian population confirmed that physical activity moderates the relationship between variants in the *FTO* gene and waist circumference (Moore et al., 2012). Qi et al. (2012) and Kilpeläinen et al. (2011) replicated this finding in a sample with 12,304 and 218,166 individuals for BMI and obesity, respectively. More recently, a large study with 252,188 unrelated individuals of European ancestry confirmed that physical activity moderates the relationship between variants in the *FTO* gene and BMI (Moore et al., 2019). These are fairly robust findings, each suggesting that physical activity can moderate the effect of *FTO* on BMI and obesity. However, physical activity is endogenous as individuals choose whether and how much to exercise. Marsaux et al. (2016) show that there is no significant association between the risk alleles of the *FTO* gene and physical activity, even if the risk allele is disclosed to the participants, providing some evidence that there is no self-selection into exercise based on the risk allele. Still, studies that explore exogenous exercise changes are rare; Mason et al. (2020) use an intention-to-treat design, exploring the interaction between *FTO* and living within 1 km of physical activity facilities. They find no evidence of an interaction between the *FTO* gene and intention-to-treat for physical exercise. On the other hand, Mitchell et al. (2010) conducted a randomized exercise intervention and found some evidence of heterogeneous effects of exercise. In particular, participants with two *FTO* risk alleles experienced a larger weight loss for higher exercise levels. All in all, while there is strong evidence for a moderation of the *FTO* by exercise, it remains unclear whether physical activity can causally change the genetic effects of this genetic variant.

Research has also suggested that the influence of the *FTO* genetic variant on BMI depends on the existence of an "obesogenic" environment, i.e., an environment characterized by food abundance and physical inactivity, which have been growing in the last decades. In line with this, Rosenquist et al. (2015) found that *FTO* is associated with a greater increase in BMI for younger as compared to older individuals (defined as those born after versus before 1942) in a sample of 3,720 individuals. Similarly, Demerath et al. (2013) found an *FTO*-by-year-of-birth interaction in a small sample of approximately 1,000 individuals. Further, Taylor et al. (2011) and Vasan et al. (2012) found that *FTO* associates more strongly with obesity in urban compared to rural populations in South Asia and India. Additionally, a lack of association of *FTO* with obesity-related outcomes was observed in a Gambian rural population, a region with predominant food scarcity (Hennig et al., 2009). Together, these studies suggest that the effect of *FTO* on BMI and obesity may be moderated by both calories in (driven by food availability) as well as calories out (driven by physical activity).[10]

*Smoking*

Another much-studied outcome is smoking, where several studies have identified an association between the genes *CHRNA5* and *CHRNA3* (the proteins encoded by these genes are part of certain nicotinic acetylcholine receptors) and smoking intensity (e.g., Caporaso et al., 2009; Thorgeirsson et al., 2008; Weiss et al., 2008). This association was later replicated for the SNPs rs1051730 and rs16969968 located in the *CHRNA3* and *CHRNA5* genes, respectively (Berrettini & Doyle, 2012; Bierut, 2010; Liu et al., 2010; Tobacco and Genetics Consortium et al., 2010; Tyrrell et al., 2012; Wehby et al., 2011). Positive associations between the SNPs and lung cancer and chronic obstructive pulmonary disease were additionally found (e.g., Bierut, 2010; Thorgeirsson et al., 2008), and both polymorphisms are highly associated with nicotine dependency, with an $R^2$ of 0.8% (Thorgeirsson et al., 2008). Recent work has suggested an $R^2$ of 0.4%–0.5% of the SNP rs1051730 for women during pregnancy (Pereira et al., 2020).

In a sample of 26,241 European participants, Tyrrell et al. (2012) find an interaction between the maternal SNP rs1051730 and smoking status in explaining offspring birth weight. This suggests the SNP influences offspring birth weight through maternal smoking, since the association is nonexistent among nonsmokers. Recent studies have corroborated this finding in two large independent samples with $N = 2,841$ (Yang et al., 2020) and $N = 256,702$ (Pereira et al., 2020). Furthermore, Fletcher (2012) showed how variation in the nicotinic acetylcholine receptor (CHRNA6) moderates the influence of tobacco taxation policy on multiple measures of tobacco use, providing evidence for a gene–policy interaction.

### Other Outcomes

In addition to BMI (obesity) and smoking, there are other outcomes where SNPs have been identified with relatively large explanatory power. An example is one specific genetic variant that explains 0.4% of the variance in human height (Lango Allen et al., 2010; Okbay & Rietveld, 2015; Wood et al., 2014), as well as variants in the *ADH1B*, *ADH1C*, and *ALDH2* genes that are involved in alcohol (ethanol) metabolism. Nevertheless, to our knowledge, there are no studies investigating the interplay between environments and other "top SNPs." Therefore, the section "Polygenic Index–Environment Interplay" discusses studies that investigate gene–environment interplay using polygenic indices instead of individual SNPs.

## *Polygenic Index–Environment Interplay*

Compared to single genetic variants, polygenic indices have substantially greater predictive power, sometimes rivaling the predictive power of traditional social science variables such as family socioeconomic status (see, e.g., Lee et al., 2018). Because of this, most recent studies exploring gene–environment interplay use polygenic indices. One important limitation of these indices is that, currently, they tend to be most predictive for populations of European ancestry (Duncan et al., 2019; Martin et al., 2019).[11]

The three subsections of this section each investigate correlations between polygenic indices and outcomes as a function of the environmental context. The first focuses on measures of the environment that are either predetermined or less prone to self-selection, such as year of birth. These type of studies usually document patterns rather than effects or causality. Although these studies cannot pinpoint the exact environmental cause of heterogeneity in the effect size of the polygenic index, they can give important clues as to how environmental factors can affect genetic influences. The second subsection reviews studies that investigate how family socioeconomic status (SES) modifies the associations between polygenic indices and outcomes. These studies are not causal, as family SES is not exogenous, but it is predetermined and usually free from self-selection issues, since children cannot choose their families. The third subsection looks at studies that investigate *exogenous* environmental changes within a framework exploring gene–environment interplay. These types of studies usually explore policy reforms, interventions, or natural experiments.[12]

Each subsection highlights work that additionally focuses on isolating exogenous variation in the genetic measure $G$. While $G$ is always predetermined—our minimum criterion for our environmental measure $E$—it is not necessarily exogenous. In fact, $G$ is exogenous only in a within-family analysis (as conditional on parental genetic makeup, the child's genetic makeup is as good as random). Indeed, comparing across families, the PGIs may be systematically correlated with an individual's ancestry, culture, and family upbringing and, with that, may capture *genetic nurture* (see, e.g., Koellinger & Harden, 2018; Kong et al., 2018). To isolate the causal (or direct) effect of the PGI, authors have used family designs (e.g., Selzam et al., 2019), sibling fixed effects (e.g., Cawley et al., 2019), adoptees (e.g., Cheesman et al., 2020), offspring phenotypic information (Wu et al., 2021), and, more recently, within-sibling GWASs (e.g., Howe et al., 2021).

## *Predetermined Environments in PGI × E*

Theoretical discussions of gene–environment interplay usually consider the most general definition of "the environment": a broad, multilevel, and multidimensional concept including all influences other than genetic inheritance (Spinath, 2010). This definition of the environment includes anything that happens "outside of the skin": from prenatal events, to nutritional deficiencies, to family and peer socialization, and the institutional and policy environment.

A precise measure of such a generic concept of the environment is often hard to come by. A common way to circumvent this problem is to use year of birth as a proxy. This type of proxy ensures that there is no self-selection based on the polygenic index or other unobserved characteristics. Unlike other context-dependent measures such as neighbourhood quality, year of birth cannot be influenced by the individuals themselves. A limitation of these studies, however, is that it is hard to pinpoint which specific feature of the environment lies at the heart of the gene–environment interplay.

Using the Health and Retirement Study (HRS), Conley et al. (2016) found that the predictive power of the educational attainment polygenic index has decreased for younger cohorts. Consistent with this result, Herd et al. (2019) documented a decrease in the predictive power of the educational attainment polygenic index for younger cohorts using three U.S. data sets (HRS, Add Health, and the Wisconsin Longitudinal Study). Additionally, Herd et al. (2019) found that the predictive power changed differentially by gender: the predictive power of the PGIs is weaker for women than for men in the 1950s and 1960s, but as constraints for women to stay in education decreased, gender differences diminished. Using the same data set (HRS), however, Lin (2020) found that the predictive power of the educational attainment polygenic index has increased (rather than decreased) for younger cohorts (born in 1942–1959) compared to older cohorts (born in 1920–1941). This inconsistency might be explained by the fact that Lin (2020) controls for parental SES, while Herd et al. (2019) and Conley et al. (2016) do not. In this sense, it is possible that the effect size of the raw polygenic index of education has increased but the effect size conditional on SES has decreased. Clearly, more research on the predictive power of the educational attainment polygenic index is necessary to draw further conclusions.

Other studies have focused on the association between polygenic indices and health outcomes, such as BMI. Conley et al. (2016) document an increase in the predictive power of the BMI polygenic index for younger cohorts in the United States, consistent with the findings of Liu and Guo (2015) and Guo et al. (2015), which suggests an interplay between the PGI and year of birth.[13]

Besides these two clusters of research on the education and BMI polygenic indices, Conley et al. (2016) find an increased predictive power of the height polygenic index, a decreased predictive power of the polygenic index for heart disease, and a stable predictive power for the depression polygenic index. Furthermore, Domingue et al. (2016) documents an increasing predictive power of the smoking initiation polygenic index, particularly for the cohorts from the 1950s and onwards.

Finally, characterizing individuals' early life circumstances by the infant mortality rate in one's year and local geographic area of birth, Baker et al. (2022) start by replicating a seminal paper by Barker and Osmond (1986), showing a strong positive relationship between infant mortality rate at birth and ischaemic heart disease in later life. In addition to exploring whether this relationship holds within local geographic areas and within families, they explore whether the positive association conceals underlying genetic heterogeneity, using one's PGI for heart disease. The findings suggest that the effects of one's early life environment and one's genetic predisposition reinforce each other. The interaction is robust to local geographic area fixed effects as well as family fixed effects. The latter family fixed effects specification estimates the direct (causal) effect of genetic variation, interacted with a predetermined environment. Their results show that, in areas with the lowest infant mortality rates, the effect of one's genetic predisposition effectively vanishes.

*Socioeconomic Status in $PGI \times E$*

Researchers interested in understanding the sources of socioeconomic inequality often compare differences in outcomes of children raised by families of different socioeconomic status (SES). Whether the relationship between family SES and outcomes changes with the genetic endowment of children has been a topic of debate, starting from Scarr-Salapatek (1971) to a more recent meta-analysis by Tucker-Drob and Bates (2016).

One strand of the literature has investigated the relationship between polygenic indices and outcomes related to economic success and social status, such as educational attainment or income. For example, again using the HRS data, Papageorge and Thom (2020) find a strong positive interaction between the educational attainment polygenic index and childhood SES in explaining higher education (college or more). This suggests that the educational attainment polygenic index is more predictive for individuals who grew up in higher SES households. On the other hand, they find a negative PGI-SES interaction when predicting high school graduation. Consistent with this finding, Pereira (2021) found no evidence of an interaction between the educational attainment PGI and childhood SES for total years of education in the HRS. Using the same data set, Lin (2020) found a negative interaction between a different measure of family SES and the educational attainment polygenic index for the probability of advancing to the next educational level, suggesting that the genetic effect becomes smaller when parental education increases. Using a within-family design to isolate the causal effect of $G$, Ronda (2020) found a positive interaction between family advantage and the educational attainment PGI for human capital formation as measured by a composite measure of attainment and grades, in a Danish data set. Belsky et al. (2018) explored five longitudinal studies in the United States, Britain, and New Zealand, comprising over 20,000 individuals in total. The authors found no evidence of a significant interaction in any of the cohorts but one (Add Health) in which they find a positive interaction. Consistent with this, Selzam et al. (2017) and Allegrini et al. (2020) found no evidence of an interaction between the educational attainment polygenic index and family SES for educational achievement or general cognitive ability in two U.K.-based cohorts. Overall, these studies do not suggest there is a robust statistical interaction between the educational attainment polygenic index and early life SES.[14]

Finally, some studies have focused on the interplay between genetic variation and individual's SES in explaining health outcomes. Liu and Guo (2015) found a positive gene–environment interaction coefficient between SES or downward mobility and the BMI polygenic index in explaining obesity, suggesting that the genetic risks of high BMI are amplified for individuals from lower SES or with downwards mobility. Treur et al. (2018) showed that the polygenic index for smoking intensity is associated with smoking intensity only in individuals exposed to parental smoking during childhood. The authors found no evidence for a gene–environment interaction for the polygenic index of smoking initiation. These results are in line with a study that suggests that childhood SES moderates the genetic risk for smoking intensity, such that having a high childhood SES protects against such genetic risk (Bierut et al., 2018).

*Exogenous Environments in $PGI \times E$*

The focus here is on more specific, albeit narrower, definitions of the environment. In particular, we consider the literature that analyzes gene–environment interplay in the presence of (quasi)-exogenous variation in policies or programs, or due to natural experiments. This is the cleanest way to identify the environmental effect, ensuring that the environment is orthogonal to any observable as well as unobservable characteristics (Biroli et al., 2021). Combined with a within-family design, this would give estimates of gene–environment interplay where *both* the genetic and the environmental effect can be interpreted as causal.

An increasing number of studies explore whether exogenous changes to the environment modify the association between the educational attainment polygenic index and educational outcomes. One of the first examples in the literature is Schmitz and Conley (2017), which explored the Vietnam lottery draft and found that veterans with below-average polygenic indices for educational attainment completed fewer years of schooling than did comparable nonveterans. No difference was found for veterans with above-average polygenic indices, suggesting that having an above-average polygenic index may have protective effects against an adverse shock, such as being drafted. Muslimova et al. (2020) also provide evidence for the existence of modifiable genetic effects. The authors explored birth order as a proxy for parental investments using a within-family design aimed at isolating the direct (causal) effect of $G$, to conclude that those who randomly inherited higher genetic endowments for education benefit disproportionally more from the additional parental investments that are associated with being firstborn compared to those with lower genetic endowments. Furthermore, Rimfeld et al. (2018) exploited a natural experiment that shifted the economic and political system from communism to capitalism. Specifically, the authors compared the predictive power of polygenic indices in pre- and post-Soviet era Estonia, finding that the educational attainment polygenic index explains twice as much variance in the corresponding outcomes in the post-Soviet era. Finally, Von Hinke and Sørensen (2022) investigated whether one's genetic predisposition to human capital and health outcomes moderates the effect of early life exposure to a severe pollution event: the London smog of 1952. They found that prenatal and early childhood exposure to the London smog reduces fluid intelligence and increases the likelihood of being diagnosed with respiratory disease later in life, and that these effects are stronger for those more genetically predisposed to those outcomes.

Other studies examine whether exogenous changes in the environment may have modified genetic influences on health outcomes. For example, Schmitz and Conley (2016) explored the Vietnam-era draft lottery and its interaction with the smoking initiation polygenic index. Using the military draft as an instrument for being a veteran, the authors found that on average, being a veteran does not increase the chances of smoking or the number of cigarettes. Yet, being a veteran with an above-average polygenic index for smoking initiation does increase the chances of smoking, having cancer, and having higher blood pressure. These results again highlight the potential of using polygenic indices for heterogeneity analysis. The authors also found that attending college can moderate the genetic risk of smoking. Similar to Fletcher (2012), but using polygenic indices instead of a specific SNP, Slob and Rietveld (2021) explored the effect of quasi-exogenous variation in tobacco excise taxation in the United States on smoking behavior (see also Fontana, 2015). They found a significant gene–environment interaction coefficient between state-level changes in tobacco taxes and two polygenic indices—the smoking initiation and smoking intensity polygenic index—but no interaction with the smoking cessation polygenic index. Their results suggest that lower taxes drive individuals with a higher genetic propensity to start smoking and to smoke more heavily.

Schmitz et al. (2021) investigated the effect of business closures and subsequent layoffs on the BMI of late career workers. The authors determined that being laid off significantly increases BMI, yet they found no significant interaction between the BMI polygenic index and being laid off. Barcellos et al. (2018) used a compulsory schooling age reform in the United Kingdom to study gene–environment interplay on BMI. The authors found a significant interaction between the BMI polygenic index and having received more education due to the reform, with the additional education generated by the policy having a protective effect on individuals with a high genetic risk of obesity.

Looking at health behaviors as the outcome of interest, Biroli and Zünd (2021) leveraged quasi-exogenous variation in alcohol licensing policies in the United Kingdom to show that the genetic predisposition to drink alcohol contributes to health inequalities in two ways: through selection into neighborhood with greater alcohol availability (i.e., gene–environment correlation), and by decreasing susceptibility to more restrictive licensing policies. Both negative selection and decreased susceptibility are related to higher alcohol intake and, eventually, alcohol-related diseases. Using minimum drinking age laws to study binge drinking, Fletcher and Lu (2021) found that the moment of passing the minimum drinking age is associated with recent binge-drinking episodes. However, those effects are entirely concentrated in individuals with a high polygenic index for alcohol use. The authors found no evidence of interaction effects between minimum drinking age laws and the risk-taking polygenic index.

Further, Biroli and Zwyssig (2021) used polygenic indices related to smoking to understand heterogeneity in morally hazardous behaviors, finding that individuals who suffer a health shock when uninsured are more likely to reduce smoking, but only if they have a low polygenic index for smoking. Finally, Barban et al. (2021) explored variation in access of the contraceptive pill and a within-family design to isolate the direct (causal) effect of $G$, and to estimate interactions between access to the pill and the polygenic indices for age at first sex, completed fertility and childlessness for the same outcomes respectively. The authors find that an anticipation of sexual debut and the postponement of motherhood led by the diffusion of the pill are magnified by gene–environment interactions at the same time that the decline in family size and the rise in childlessness associated with the diffusion of the pill are attenuated by gene–environment interactions.

## Policy Implications

There are many potential benefits of social science genetic research (see, e.g., Harden, 2021b). This chapter focuses on studies that examine gene–environment interplay, and we argue that a better understanding of the interplay is interesting from an academic perspective, but also relevant to policy. Indeed, it is increasingly accepted that the effect of genetic variation on behavior, health, or economic outcomes often depends on environmental exposures and vice versa. Understanding such environmental circumstances and critical periods of life that affect human capabilities and unhealthy behaviors for different genetic risk groups can help in shaping policy.

Indeed, *because* the genetic makeup of individuals does not change over the life course, policy makers can focus on how *environmental* changes may differentially affect individuals with high or low values for polygenic indices. Evidence of such heterogeneity in the responses to certain interventions may help shape government policy, as it is informative about whether certain policies increase or reduce genetic inequalities in the population. For example, the main policy tools credited with the significant reduction in the rate of smoking are cigarette taxation and information dissemination (Flor et al., 2021; Golechha, 2016). Yet, many smokers seem resistant to the existing policy repertoire. Why do they not respond while others do? Genetic information can provide insight into unobserved heterogeneity in policy responses, which can be used to improve policy (Fletcher, 2012; Slob & Rietveld, 2021). If those who keep smoking are solely genetically at-risk individuals, one might be limited to pharmacotherapy or pharmacogenetics. But if they are genetically at-risk individuals who have experienced adverse (childhood) environments, prevention efforts targeting modifiable characteristics of such environments may reduce later-life dependence. Social science genetic research therefore can point to means of intervention.

Clearly, we should beware of government policies targeting groups based on genetic information. But the policy solutions suggested in this article do not require that policy makers know the genetic makeup of individuals in the population, just the existence (or not) of a differential response based on the genotype. To obtain this knowledge, only the researcher needs to have access to the anonymized genetic information, which must be handled according to the carefully drafted ethical, privacy, and confidentiality guidelines currently in place, especially given the hideous history of the use of genetic data in social science in the early 1900s. Indeed, most researchers working in this field are concerned about refueling similar thinking by studying human genetics. However, the solution to this is not simply to ignore the existence of genetic differences across human beings, but rather to acknowledge it (Harden, 2021b).

A better understanding of gene–environment interplay may also lead to personalized approaches, in which genetic information is used to inform individuals to avoid environments that may harm them. For example, in personalized medicine, genetic information on individuals' risk of developing a disease can be used to decide whether to embark on a particular (preventive) treatment (see, e.g., Khera et al., 2018; Torkamani et al., 2018). Personalized education, on the other hand, is more controversial. A common argument against targeting education policies based on pupils' genetic predispositions is that the polygenic indices that are available at the moment are simply not sufficiently accurate to obtain robust individual-level predictions (Morris et al., 2020). But what *if* we get to a point where individual-level predictions are sufficiently accurate? Could personalized education be part of a package of policies to help everyone achieve their full potential? While this might sound like an appealing idea at first, caution must be taken. The current measures of potential based on genetic variants are extremely narrow and centered around individuals of European ancestry. These genetic measures also depend highly on the context in which they were obtained (western educated industrialized rich and democratic [WEIRD] populations and environments). Further, people enjoy different talents that are not always reflected in academic achievement or cognition. In this sense, trying to optimize students to fit the schooling system might arguably be worse than trying to optimize the schooling system so that everyone can grow their unique talents. That said, this discussion should be carried out in tandem with other social scientists, ethicists, and interest groups that might be directly affected by such policies.

Finally, the findings from studies investigating gene–environment interplay might also inform policies that aim to promote equality of opportunity. Such policies try to limit the extent to which "circumstances" affect one's outcomes. If genes can be seen as circumstances, [15] policy makers can have a better picture of the true responsibility (or lack thereof) that individuals have for their outcomes. For that reason, an important takeaway from the gene–environment literature is that the two components of "nature" and "nurture" *jointly* shape individuals' outcomes. With that, it provides a strong argument against genetic or environmental determinism; the belief that individuals' outcomes are shaped only by genetic variation or only by environmental conditions, with no room for alternative mechanisms (see also, e.g., Caspi et al., 2010; Manuck & McCaffery, 2014). This research therefore has the potential to renew the debate on equality of opportunity, informed by new evidence on the role of genes, environments, and their interaction.

## Open Questions and Research Avenues

Research at the intersection between genetics and the social sciences is still in its infancy. However, the field has progressed incredibly quickly in recent years, and there is no evidence of it slowing down. If anything, it seems to be taking off, with efforts in data collection such as the "All of Us" study, which seeks to collect whole genome data on more than a million Americans, as well as other large data collection efforts such as the "1+ Million Genomes Initiative." With rapid growth in sample size, there is an increasing focus on causal genetic inference, made possible by growing efforts in family-based genetic discovery (see, e.g., Howe et al., 2021) and family-based gene–environment analyses (see, e.g., Muslimova et al., 2020). This section concludes by looking forward with a discussion of a few important avenues for future development and application.

## *Interplay between Nature and Nurture*

The interplay between nature and nurture in shaping economic outcomes is likely to stay relevant for decades to come. Widespread access to polygenic indices is providing researchers with a great novel tool to have a continuous proxy of "nature," where historically these measures were implicit (e.g., in twin or adoption studies) or contaminated by environmental influences (e.g., birth weight, or early life test scores). Early examples of actual nature-nurture interplay estimation have been reviewed in this chapter, but a full understanding of the interplay is still in its infancy. For researchers interested in causal studies of gene–environment interplay in the social sciences and economics, there are several important issues to consider. First, although genetic data is increasingly included in data sets used by social scientists, most data sources do not collect participants' DNA. As we have argued, there are substantial returns to the collection of genetic data. Second, to obtain causal estimates of the genetic component, it is crucial to observe genetic data on multiple members of a family (e.g., siblings or parent–child trios). Although this is still relatively rare, these data are increasingly being collected. Third, to obtain causal estimates of the environmental component, it is necessary to have some (quasi-)random variation in the environment of interest. Combining this with a family-design would require one sibling to be exposed to the treatment of interest, while the other is not. Again, this is rare. This may be why the literature so far had to be a little opportunistic, exploiting exogenous shocks that happen to be available in data sets that include genetic data. We work with what we have. Until more data become available, this is likely to drive the literature. These studies nonetheless have increased our understanding of the importance of gene–environment interplay in shaping individuals' outcomes.

Moving forward, with the increasing availability of genetic data, reporting heterogeneity of treatment effects by genetic characteristics might become a routine, just as is currently done for other predetermined factors such as age and sex. An increased used of genetic data in randomized controlled trials and treatment evaluations not only allows researchers to discover interesting and previously hidden sources of heterogeneity, but can also shed light on inequality of opportunity and intergenerational mobility.

## *Inequality of Opportunity and Intergenerational Mobility*

Genetic data, and its interplay with socioeconomic indicators, has the potential to greatly enrich existing estimates of inequality of opportunity and intergenerational mobility. In both fields of study, the focus is currently solely on socioeconomic barriers to equality and social mobility. For example, in inequality of opportunity studies, a typical application includes a rich set of socioeconomic background characteristics (e.g., parental education, parental occupation, parental income, race, sex, and residential area of origin) to analyze inequalities in education or income with respect to this set of circumstances, for which individuals cannot be held responsible (see, e.g., Roemer & Trannoy, 2016). Intergenerational mobility typically adopts a similar perspective but instead focuses on one of these "circumstances" and directly computes the association between the parental and offspring outcome of interest.

In intergenerational mobility research, genetic data can be used to explore the extent to which intergenerational persistence of a trait has both environmental and genetic origins (e.g., Isungset et al., 2021). Previously, scholars had to rely on rare cases where adoption procedures were of a random nature (e.g., Fagereng et al., 2021; Sacerdote, 2007). In the inequality of opportunity framework, the existence of "innate ability" or "talents" has long been acknowledged (e.g., Rawls, 1971). Genetic data can be used to estimate the share of "genetic advantage" Pereira (2021) or "genetic lottery" (Harden, 2021b) more directly. While it is unclear whether this genetic component should be perceived as a fair or unfair source of advantage, it is clear that genetic components matter for the realization of outcomes that society considers important. If one adopts the view that genetic components are unfair sources of advantage, then genetic components and their interactions with socioeconomic indicators could all be included in the share of inequality of opportunity (e.g., Pereira et al., 2020).

## *Understanding Human Capital Formation*

Another strand of economic literature that can benefit from the use of genetic data is the study of the human capital formation. A number of theories involve endogenous responses and interactions with what is often denoted as "endowments." For example, Becker and Tomes (1976) explored theoretically the parental investment response to differences in the child's endowments. Cunha and Heckman (2007) developed a model of skill formation where endowments and parental investments are complementary, a concept known as dynamic complementarity. Genetic data, and in particular polygenic indices, have the potential to be useful measures of "endowments" and thus help shed new empirical insights on these theoretical hypotheses. Indeed, Breinholt and Conley (2020) provide evidence supporting Becker and

Tomes (1976), finding that parents provide more cognitive stimulation to children with a higher education PGI. Another example is the work of Muslimova et al. (2020), where interactions between polygenic indices and birth order (used as a proxy for parental investments, following the birth order literature) were investigated as an empirical test of dynamic complementarity. Finally, Houmark et al. (2020) incorporated genetic endowments into a dynamic latent factor model previously proposed by Cunha and Heckman (2007), allowing them to identify different genetic mechanisms and to control for measurement error in skills and investments.

### *Improved Measurement of Genetic Predispositions in Diverse Populations*

Improving the predictive power of polygenic indices has been a main focus in past years (Ge et al., 2019; Privé et al., 2021; Vilhjalmsson et al., 2015) and has been very successful. More recent literature, however, highlights important issues in interpreting what exactly these polygenic indices capture (see, e.g., Kong et al., 2018), whether the construction method matters (Muslimova et al., 2020), and how to deal with measurement error in polygenic indices (Becker et al., 2021; Van Kippersluis et al., 2021).

One of the largest drawbacks of current social science genetics research is that polygenic indices are calibrated for populations of European ancestry. Martin et al. (2019) estimated that 79% of participants in genetic studies are of European descent, despite making up only 16% of the global population. One implication of this lack of representation is that polygenic indices have less predictive power in non-European populations (Wojcik et al., 2019). This is consequential, since any benefits of genetically based research are then likely to be concentrated among those of European ancestry, at the expense of a better understanding among underrepresented populations, potentially leading to increased inequalities. One important future avenue of research is therefore to conduct GWASs on non-European ancestry samples. Indeed, studies that include diverse populations (Shelton et al., 2021; Vargas et al., 2016; Wojcik et al., 2019) including Asian (Cai et al., 2015), African (Duncan et al., 2018; Gerhard et al., 2018; Wang et al., 2020), Hispanic (Qi et al., 2017), and Indigenous populations are crucial for a better understanding of human genetics and how the environment may moderate the effects of genetic variation in different populations.

Another avenue for future research is to explore whether particular SNPs affect not only the mean of an outcome, but also other moments, such as the variance. Current GWAS are optimized to measure the association between a SNP and the mean outcome, but exploring the relationship with the variance of the outcome (Schmitz et al., 2021), or the putative biological pathways of the relationship (Hu et al., 2017), may be similarly fruitful. Such associations can then be taken into account when constructing different polygenic indices, and they may be able to capture different types of gene–environment interplay.

Last, an exciting new development has been a recent focus on family-based GWASs and family-based gene–environment analyses. Currently, because genes and environments are not independent (e.g., a person inheriting genetic variants associated with nicotine dependence also is more likely to have grown up with smoking parents), genetic variants partly reflect individuals' environments. This may lead to false identification in genetic discovery studies (GWASs) of genetic variants that in reality bear no causal relation to those outcomes. Fortunately, a person's genetic makeup is a random combination of those of his/her parents (Mendel's First Law). Therefore, by observing both parental and offspring genotypes (so-called

parent–child dyads), it is possible to identify genetic variants that are truly causally implicated in human traits and diseases. This holds true also for studies that include genetic data on siblings. A recent meta-analysis on 160,000 siblings from 17 cohort studies demonstrated the importance of family-based GWASs, in particular for behavioral and socioeconomic traits (Howe et al., 2021), such as those discussed here. It is anticipated that genetic discovery results from increasingly large family-based GWAS consortia will become available in the next decades. In addition, it is foreseen that family-based data sets for gene–environment analyses will also increasingly become available. This will enable analyses of the causal effects of genetic variants and polygenic indices in gene–environment interplay. Combined with exogenous variation in environments, such analyses would constitute the ideal experiment.

In short, we are at the start of what an ever-growing field of research, made possible by rapid advances in methods and the availability of genetic data. These innovations will help us answer age-old questions that are of interest to economists and other social scientists. It is not too late to join the bandwagon. If not you, who? If not now, when?

## Further Reading

*Essential Reading*

Conley, D. C., & Fletcher, J. M. (2017). *The genome factor: What the social genomics revolution reveals about ourselves, our history, and the future*. Princeton University Press.

Freese, J. (2018). The arrival of social science genomics. *Contemporary Sociology: A Journal of Reviews*, *47*(5), 524–536.

Harden, K. P., & Koellinger, P. D. (2020). Using genetics for social science. *Nature Human Behaviour*, *4*(6), 567–576.

Martschenko, D. O., & Smith, M. (2021). Genes do not operate in a vacuum, and neither should our research. *Nature Genetics*, *53*(3), 255–256.

*Nonessential Reading*

Tam, V., Patel, N., Turcotte, M., Bossé, Y., Paré, G., & Meyre, D. (2019). Benefits and limitations of genome-wide association studies. *Nature Reviews Genetics*, *20*(8), 467–484.

Trenkmann, M. (2018). Lessons from 1 million genomes. *Nature Reviews Genetics*, *19*(10), 592–593.

Rita Dias Pereira, Pietro Biroli, Titus Galama, Stephanie von Hinke, Hans van Kippersluis, Cornelius A. Rietveld, and Kevin Thom

*Figure 1*. A single nucleotide polymorphism (SNP).

---

[1] "The difference of natural talents in different men is, in reality, much less than we are aware of; and the very different genius which appears to distinguish men of different professions, when grown up to maturity, is not upon many occasions so much the cause, as the effect of division of labour. The difference between the most dissimilar characters, between a philosopher and a common street porter, for example, seems to arise not so much from nature, as from habit, custom, and education. When they come into the world, and for the first 6 or 8 years of their existence, they were, perhaps, very much alike, and neither their parents nor play-fellows could perceive any remarkable difference" (Smith, 1776, pp. 28–29).

² For a brief history of the literature incorporating genetic concepts into economic analysis, see, e.g., Fletcher (2018).

³ In early intergenerational mobility studies, the genetic channel is typically quantified using kinship correlations (e.g., Becker, 1976; Goldberger, 1979), siblings (e.g., Behrman & Wolfe, 1989; Griliches, 1979), adoptees (Plug, 2004; Sacerdote, 2002; Wilcox-Gök, 1983) or twins (e.g., Behrman & Taubman, 1989; Taubman, 1976).

⁴ The control variables typically include age, sex, and the first set of (e.g., 10) principal components of the genome-wide matrix of all SNPs to control for population stratification (Price et al., 2006). Population stratification is a form of confounding, where ancestry influences one's genetic makeup as well as the outcome $y_i$ through nongenetic pathways, such as culture. More specifically, if a population is stratified into subpopulations that do not mate randomly, and an outcome happens to be more common in one subpopulation for nongenetic (e.g., cultural) reasons, then the outcome will appear to be correlated with any SNPs that also happen to be more common in that subpopulation. Population stratification reflects systematic differences in allele frequencies between different ethnic groups that may arise from nonrandom mating or simply through random drift as populations are segregated (e.g., due to natural barriers such as rivers and mountains or distance). A failure to account for this may lead to false associations between the genotype and the outcome of interest (Hellwege et al., 2017). A commonly used hypothetical example is the "chopstick gene," where people of Asian descent have different allele frequencies and tend to eat with chopsticks for cultural reasons. A GWAS, investigating the genetic basis of chopstick use, without controlling for ancestral differences in allele frequencies, would then pick up "chopstick genes." The standard way to correct for population stratification is to restrict the analyses to people who have similar genetic ancestry and to control for principal components that capture broad patterns of genomic similarity produced by human demographic history (Price et al., 2006). Nonetheless, recent studies suggest that migration may have

resulted in a geographical clustering of common genetic variants that cannot be adequately captured by principal components (e.g., Abdellaoui et al., 2019).

[5] Individual SNPs tend to be insufficiently predictive to be used on their own, though there are some exceptions, such as the FTO gene for obesity, the $\alpha 5$ nicotinic receptor subunit gene, and the $\alpha 6 \beta 3$ nicotinic receptor subunit genes for nicotine dependence, the ADH1B gene for alcohol consumption, or the APOE $\varepsilon$-4 allele for Alzheimer. Some of these are discussed in the context of gene–environment interplay (see the section "Candidate Gene Studies").

[6] In some settings, researchers use only genome-wide significant SNPs to create the PGI. Other methods try to account for the correlation structure of SNPs—so-called linkage disequilibrium (LD)—in the human genome. This is done because LD may lead to "double-counting" of genetic regions, especially those with strong effects (see for example, So & Sham, 2017; Vilhjalmsson et al., 2015). An example of such methods is "clumping," which uses only the most significant SNPs in a small, highly correlated region. More sophisticated ways to account for LD exist, such as a Bayesian re-sampling of weights based on an external reference panel with LD information (see, for example, Privé et al., 2021; Vilhjalmsson et al., 2015).

[7] Another advantage of a within-family analysis is that, given that genetic variation is randomized across siblings, the coefficient on the genetic term is free from population stratification.

[8] The empirical study of gene–environment interplay started by using twin and adoptee samples. For an early appraisal of the literature, see Plomin et al. (1988). This literature is now vast and spans several disciplines, including economics, where the focus has often been on understanding the interplay in the nature and nurture of skills, education, wages, and income (see, among others, Black et al., 2020; Björklund et al., 2005, 2006, 2007; Brandén et al., 2017). A recent effort to improve the statistical power, validity, and representativeness of these family-bases studies culminated in the establishment of the Consortium on Interplay of Genes and Environment Across Multiple Studies (IGEMS) (Pedersen

et al., 2019). A novel approach using in vitro fertilization and donor children instead of twins or adoptees has been suggested by Lundborg et al. (2021).

[9] The SNP rs9939609 is one of 10 SNPs in the first intron of the FTO gene. All BMI-associated SNPs are highly correlated with each other ($R^2$ from 0.5 to 1.0). The SNP rs9939609 was chosen by the authors because it has the highest genotyping success rate (100%) among the cluster of most highly associated SNPs.

[10] Many other studies have suggested possible interactions with diet (e.g., Lappalainen et al., 2012), though its results were not replicated in two recent meta-analyses (Nettleton et al., 2015; Qi et al., 2014).

[11] The research community is acutely aware of these limitations and the potential unfair consequences of research being conducted mostly on populations of European ancestry. Efforts and resources are directed to address this disparity, as discussed in the section "Open Questions and Research Avenues".

[12] This review therefore omits studies that use endogenous measures of the environment. Although such work is important and can be used to identify potential causal pathways, the interpretation of the gene–environment interaction coefficient is not straightforward if the environment is endogenous (see also Biroli et al., 2021).

[13] Domingue et al. (2020) argue that part of the increase in the predictive power of the BMI polygenic index for younger cohorts, especially for men, is due to the increasing variance in BMI over time.

[14] Other studies have focused on the interaction between the educational attainment polygenic index and the average socioeconomic status of the school attended by the child (Harden et al., 2020; Trejo et al., 2018). These are not discussed here, since children with higher polygenic indices for education may be more likely to self-select into schools with different average quality, which complicates the interpretation of the estimate.

[15] Note that this point is far from being agreed on. Some authors argue that natural talents should receive compensation according to the equality of opportunity framework (Nozick, 1974; Rawls, 1971).